\documentclass[11pt,draftclsnofoot,onecolumn]{IEEEtran}

\usepackage[cmex10]{amsmath}
\usepackage{amssymb}

\usepackage{graphicx}

\newtheorem{defi}{Definition}
\newtheorem{theorem}{Theorem}

\newtheorem{remark}{Remark}
\newtheorem{corollary}{Corollary}

\newcommand{\Xc}{\mathcal{X}}
\newcommand{\Yc}{\mathcal{Y}}
\newcommand{\Cc}{\mathcal{C}}
\newcommand{\Mc}{\mathcal{M}}
\newcommand{\Rc}{\mathcal{R}}
\newcommand{\Ec}{\mathcal{E}}
\newcommand{\Pc}{\mathcal{P}}
\newcommand{\Nc}{\mathcal{N}}
\newcommand{\Tc}{\mathcal{T}}
\newcommand{\Sc}{\mathcal{S}}
\newcommand{\Bc}{\mathcal{B}}

\newcommand{\Vc}{\mathcal{V}}

\newcommand{\mysmallarraydecl}{\renewcommand{%
\IEEEeqnarraymathstyle}{\scriptscriptstyle}%
\renewcommand{\IEEEeqnarraytextstyle}{\scriptsize}%
\renewcommand{\baselinestretch}{1.1}%
\settowidth{\normalbaselineskip}{\scriptsize
\hspace{\baselinestretch\baselineskip}}%
\setlength{\baselineskip}{\normalbaselineskip}%
\setlength{\jot}{0.25\normalbaselineskip}%
\setlength{\arraycolsep}{2pt}}

\begin{document}
%
\title{Active Adversaries from an Information-Theoretic Perspective:\\Data Modification Attacks}
%
%

\author
{
Mahtab Mirmohseni and Panagiotis Papadimitratos\\
KTH Royal Institute of Technology, Stockholm, Sweden \\
Email: \{mahtabmi,papadim\}@kth.se}

\maketitle

\begin{abstract}
We investigate the problem of reliable communication in the presence of active adversaries that can tamper with the transmitted data. We consider a legitimate transmitter-receiver  pair connected over multiple communication paths (routes). We propose two new models of adversary, a ``memoryless'' and a ``foreseer'' adversary. For both models, the adversaries are placing themselves arbitrarily on the routes, keeping their placement fixed throughout the transmission block. This placement may or may not be known to the transmitter. The adversaries can choose their best modification strategy to increase the error at the legitimate receiver, subject to a maximum \emph{distortion} constraint. We investigate the communication rates that can be achieved in the presence of the two types of adversaries and the channel (benign) stochastic behavior. For memoryless adversaries, the capacity is derived. Our method is to use the typical set of the anticipated received signal for all possible adversarial strategies (including their best one) in a compound channel that also captures adversarial placement. For the foreseer adversaries, which have enhanced observation capabilities compared to the memoryless ones, we propose a new coding scheme to \emph{guarantee} resilience, i.e., recovery of the codeword independently of the adversarial (best) choice. We derive an achievable rate and we propose an upper bound on the capacity. We evaluate our general results for specific cases (e.g., binary symbol replacement or erasing attacks), to gain insights.
\end{abstract}



%
\IEEEpeerreviewmaketitle


\section{Introduction}
Operation in adverse networks requires secure and reliable communication: data modifications should not be merely detected but data should be delivered (decoded correctly) at their destination. Cryptographic primitives can ensure detection but not correction and thus data delivery. Consider a general network connecting a Transmitter (Tx) - Receiver (Rx) pair over multiple disjoint communication paths (e.g., multiple frequency bands or antennas in wireless networks, or multiple routes in multi-hop networks); adversaries can be present in a number of those paths. The challenge is how to leverage the available alternative paths in order to achieve reliable communication in the presence of the adversary. What is the best one can do against a powerful adversary? More generally, what is the best communication rate one can achieve in the face of malicious faults (adversarial modifications) \emph{and} benign faults (due to the communication channel stochastic behavior)?

Facets of this problem were addressed in the literature. One approach leverages cryptographic primitives to detect modifications and attempt retransmissions over alternative communication paths (while introducing redundancy to tolerate faults) \cite{PapHaaJSAC06}. This, however, does not address the fundamental limits of the system performance. Without cryptographic assumptions, the minimum needed connectivity is derived for resilient communication for a Tx-Rx pair over $n$ disjoint paths, termed \emph{wires}, and disrupted by active adversaries that compromise a subset of these wires (the scenario is termed the \emph{Dolev model}) \cite{Dolevetal93}. The analysis in \cite{Dolevetal93} does not consider communication rates and thus does not even attempt to achieve the best performance; it does not model channel noise and does not consider adversarial limitations or fine-grained actions.

In contrast, confidentiality received significant attention, notably after Wyner's seminal paper \cite{Wyn75}, with the majority of works concerned with passive eavesdroppers \cite[Chapter~22]{ElgKim11}. Less attention, in an information-theoretic sense, was paid to \emph{active} adversaries that modify the channel input of the legitimate transmitter. An early characteristic model is the Arbitrarily Varying Channel (AVC) \cite{BlaBreTho60}, which assumes worst-case interference: the adversary controls the channel state to maximize the error probability at the receiver. Depending on what the adversary knows and the common randomness of the legitimate nodes, the capacity can differ considerably \cite{BocSch13,Sar10}. However, it is not easy to translate erasing and replacement attacks to the AVC worst-case interference notations. In particular, AVC cannot capture data modification attacks or network structure, e.g., as the Dolev model does \cite{Dolevetal93}. Given that confidentiality (passive adversaries) is broadly researched in the information-theoretic sense (also in \cite{Dolevetal93}), \emph{the challenge is how to achieve (secure and) reliable communication in the presence of active adversaries, in addition to channel noise, and derive fundamental limits of the capacity?}

In this paper, we address this challenge. We propose a novel information-theoretic setup that captures network structure, fine-grained and strong, yet realistic active adversarial behavior, along with channel stochastic behavior. We consider a Tx-Rx pair communicating across a number of disjoint paths (routes). The adversaries compromise a fixed number of these routes, thus they get access to the respective (noiseless) transmitted signals. The adversaries can choose their best strategy (knowing the transmitted signal) to modify and increase the error at the Rx. However, their mapping is subject to a maximum distortion constraint, i.e., a \emph{distortion limit}. This limit, given a distortion measure (depending on the specific attack), determines the distance between the transmitted codeword and its modified version; e.g., for an erasing attack on binary transmissions, the percentage of bits the adversary can erase. The adversaries' placement (on the routes) is arbitrary but fixed throughout one transmission block; moreover, it may be known to the Tx. The adversaries' observations (of the transmitted signal) can be either instantaneous or cover the entire codeword. We propose accordingly two adversary types: \emph{memoryless} and \emph{foreseer}. Our goal is to find the reliable communication rate a Tx-Rx can achieve in the presence of either of these two types of adversary.

Our average distortion limit and the consideration of channel stochastic behavior (noise) on top of adversarial faults lead to a generalized model compared to the Dolev one for active adversaries. The channel noise we introduce in our model, which allows us to take into account benign faults and noisy observations, is not taken into account in the Dolev model. The distortion limit allows practical assumptions, e.g., adversaries with noisy observations, with tactics to remain undetected, limited resources or time or attempts to mount an attack, or even cryptographic integrity protection for parts of the messages (e.g., immutable fields). By setting the noise to zero and the distortion limit to its maximum, we reduce our model to the Dovel model.

We derive the capacity for the memoryless adversaries. For the achievability part, we use a compound channel to model the adversaries' placement. For each compromised route, we consider the typical set of the anticipated received signals in all possible adversarial scenarios (including the one for the best adversarial strategy), subject to the distortion limit. Then, for the foreseer adversaries, we propose a coding scheme using two techniques: (i) the Hamming approach \cite{Ham50}, to cope with worst-case errors inflicted by adversaries with access to the entire codeword, and (ii) a random coding argument, to recover from the channel stochastic noise. For the former, we use the Varshamov construction \cite{Var57}, to guarantee the required minimum distance needed to mitigate adversary-inflicted errors. Moreover, we obtain an upper bound to the capacity, taking an approach similar to that for the Hamming bound (i.e., limiting the volume of the Hamming balls). Finally, we gain insights through three special cases: replacement and erasing attacks on binary transmission and Gaussian jamming. We determine the proper distortion measures and channel distributions to model attacks that correspond to realistic situations, e.g., bit or packet replacement and dropping (selective forwarding), and evaluate our derived rates for those. For these cases, we consider explicitly the best adversarial strategy: we show the adversaries can achieve the lower bounds on the capacity we derived without specific assumptions on the adversary strategy. Our results for these special cases reveal that (i) knowing the adversaries' placement at Tx is not useful in terms of the achievable reliable rate, (ii) memory helps the adversaries significantly, and (iii) differentiates the foreseer effect from channel noise; while the memoryless effect is equivalent to channel noise.

\section{Channel Model}\label{sec:model}

\textbf{Notation}: Upper-case letters (e.g., $X$) denote Random Variables (RVs) and lower-case letters (e.g., $x$) their realizations.
$X^j_i$ indicates a sequence of RVs $(X_i,X_{i+1},...,X_j)$; we use $X^j$ instead of $X^j_1$ for brevity.
The probability mass function (p.m.f) of a RV $X$ with alphabet set $\Xc$ is denoted by $p_X(x)$; occasionally subscript $X$ is omitted.
The set of all possible distributions on $\Xc$ is denoted by $\Pc(\Xc)$. $\pi(x,y|x^n,y^n)\in\Pc(\Xc\times\Yc)$ shows the joint type (i.e., empirical p.m.f) of two sequences of length $n$, which can be extended to several $n$-length sequences. $\Pc_n(\Xc)\subset\Pc(\Xc)$ consists all possible types of sequences $x^n\in\Xc^n$. For $q\in\Pc^n(\Xc)$, the type class is defined as $\Tc^n(q)=\{x^n, p_{X}(x)=q\}$.
$A_\epsilon^n(X,Y)$ is the set of $\epsilon$-strongly, jointly typical sequences of length $n$.
$\Nc(0,\sigma^2)$ denotes a zero-mean Gaussian distribution with variance $\sigma^2$.
$\Bc(\alpha)$ is a Bernoulli distribution with parameter $\alpha\in[0,1]$.
$\mathbb{F}_{q}$ is a finite field with $q$ elements.
We define $[x]^+=\max\{x,0\}$. Unless specified, logarithms are in base 2.
Throughout the paper, $i$ and $j$ indices are used for time and route number, respectively. $H_q:[0,1]\rightarrow \mathbb{R}$ is the Hilbert $q$-ary entropy function $H_q(x) = x\log_q(q - 1) - x \log_q x - (1 - x) \log_q(1 - x)$. 
Bold letters are used to show the column vectors of length $n_r$, e.g.,
$\mathbf{x}^n= \left[\begin{IEEEeqnarraybox*}[\mysmallarraydecl]
[c]{,c/c/c,}
x_{1,1}&\ldots&x_{1,n}\\
\vdots&\ddots&\vdots\\
x_{n_r,1}&\ldots&x_{n_r,n}%
\end{IEEEeqnarraybox*}\right]$ and $\mathbf{x}^n(j)$ shows its $j$th row.

\begin{figure}[tb]
  \centering
  \includegraphics[width=10cm]{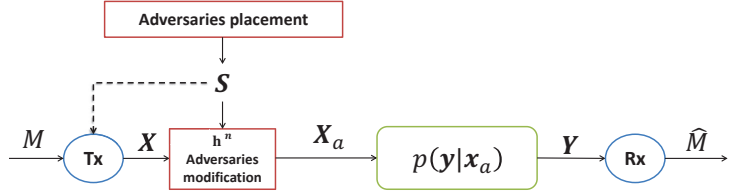}
  \caption{Multi-route Point-to-Point channel with Modifying Adversaries (PP-MA).}
  \label{fig:AADR_BD_gen}
\end{figure}
\textbf{Channel model}: Consider a single unicast scenario: Tx sends a message $M$ to Rx, with $n_r$ available disjoint routes. $n_a$ out of the $n_r$ routes are attacked by the adversaries, with their placement being arbitrary but fixed throughout one transmission block. The placement can be chosen by the adversaries to maximize the error at Rx; but, it may be known to the Tx. One can implicitly assume there are $n_a$ adversaries: more than one adversary in a route can be modeled as a stronger adversary (i.e., with a higher distortion limit). We model this scenario with a (compound) state-dependent \emph{multi-route Point-to-Point channel with Modifying Adversaries (PP-MA)} illustrated in Fig.~\ref{fig:AADR_BD_gen}: its transition probability is not entirely specified unless the \emph{Channel State Information (CSI)} (i.e., adversaries' placement information) is known \cite{CsiKor82}. Consider finite alphabets $\Xc,\Xc_a,\Yc$. The channel inputs at the Tx and the adversaries are defined by $\mathbf{X}\in\Xc^{n_r}$ and $\mathbf{X_a}\in\Xc_a^{n_r}$ respectively. $\mathbf{Y}\in\Yc^{n_r}$ is the output of the channel at Rx. The $j$-th element of state vector $\mathbf{S}\in\{0,1\}^{n_r}$, i.e., $\mathbf{S}(j)$, determines the presence of an adversary in $j$-th route. The received signal at Rx only depends on the adversary input, if present.
Each adversary channel input must be relatively close to the Tx input in that route (subject to a distortion limit), according to some distortion metric. Hence, we define the $D$ class of adversaries for $j\in\{1,\ldots,n_r\}$ by the set of all probability distributions:
\begin{align}
\Pc_j^a(D)=\{p_j(x_{a}^n|x^n): E_{p_j^n}[d(X_{a}^n,X^n)]\leq D\}\label{eqn:pmf_a}
\end{align}
where $d$ is a distortion measure defined by the mapping $d:\Xc\times\Xc_a\mapsto[0,\infty)$ and the average distortion for two sequences is $d(x_{a}^n,x^n)=\frac{1}{n}\sum\limits_{i=1}^nd(x_{a,i},x_i)$. We assume the $\Xc_a\mapsto\Yc$ channel is memoryless, thus the transition probability can be expressed by the conditional p.m.f on $\Yc\times\Xc_a$ as:
\begin{IEEEeqnarray*}{rcl}\label{eqn:pmf}
p(\mathbf{y}^n,\mathbf{x}_a^n|\mathbf{x}^n,\mathbf{s}^n)&=&p(\mathbf{y}^n|\mathbf{x}_a^n)p(\mathbf{x}_a^n|\mathbf{x}^n,\mathbf{s}^n)\yesnumber\\
&=&\prod\limits_{j=1}^{n_r}p_j(\mathbf{y}^n(j)|\mathbf{x}_a^n(j))p_j(\mathbf{x}_a^n(j)|\mathbf{x}^n(j),\mathbf{s}^n(j))\\
&=&\prod\limits_{j=1}^{n_r}p_j(\mathbf{x}_a^n(j)|\mathbf{x}^n(j),\mathbf{s}^n(j))\prod\limits_{i=1}^{n}p_j(\mathbf{y}_i(j)|\mathbf{x}_{a,i}(j))
\end{IEEEeqnarray*}
The state vector, assumed fixed in one transmission block, models the channel statistics transmission block: $\mathbf{s}_i(j)=\mathbf{s}(j)$ for $i\in\{1,\ldots,n\}$ with at most $n_a\leq n_r$ adversaries, i.e., $w_H(\mathbf{s})\leq n_a$. Hence, for $j\in\{1,\ldots,n_r\}$:
\begin{IEEEeqnarray*}{rcl}\label{eqn:pmf2}
p_{j\,X_a|X,S}(\mathbf{x}_{a}^n(j)|\mathbf{x}^n(j),\mathbf{s}^n(j))&=& p_{j\,X_{a,s}|X}(\mathbf{x}_{a}^n(j)|\mathbf{x}^n(j))=q_{j,s}(\mathbf{x}_{a}^n(j)|\mathbf{x}^n(j))
\end{IEEEeqnarray*}
where
\begin{align}
q_{j,s}(x_{a}^n|x^n)\in\Pc^a_j(D_{j,s}=s\cdot D_j)\label{eqn:pmf3}
\end{align}
which is due to the $D_j$ distortion limit at each adversary. 
In $n$ channel uses, Tx sends $M$ to Rx using the following code:
\begin{defi}\label{def:code}
A $(2^{nR},n,P_e^{(n)})$ code for the multi-route PP-MA consists of:
\begin{enumerate}
  \item A message set, $\Mc=[1:2^{nR}]$, with message $M$ uniformly distributed over $\Mc$.
  \item An encoding function, $f^{n}$, at Tx, which maps $M$ to a codeword $\mathbf{x}^n\in\Xc^{n_r\times n}$.
  \item A set of adversaries' mapping, $\mathbf{h}^{n}$, with $h^{n}(j):\Xc^{n}\times\{0,1\}\mapsto\Xc_a^{n}$ for $j\in\{1,\ldots,n_r\}$ satisfying \eqref{eqn:pmf3}.
  \item A decoding function at Rx, $g: \Yc^{n_r\times n}\mapsto\Mc$.

  \item The probability of error for this code, defined as:
    \begin{align}\label{eqn:def_Pe}
     P_e^{(n)}=\frac{1}{2^{nR}}\sum\limits_{m\in\Mc}{Pr(g(\mathbf{y}^{n})\neq m | m\textrm{ sent})}.
    \end{align}
\end{enumerate}
In case the CSI is available at the Tx, we have: $f^{n}: \Mc\times\{0,1\}^{n_r}\mapsto\Xc^{n_r\times n}$. All codewords are revealed to all nodes (including adversaries). However, the adversaries' mapping is not known to the legitimate Tx and Rx.
\end{defi}
\begin{defi}\label{def:rate}
A rate $R$ is achievable if there exists a sequence of $(2^{nR},n,P_e^{(n)})$ codes such that for $\forall\mathbf{s}\in\{0,1\}^{n_r}:w_H(\mathbf{s})\leq n_a$ and $\forall\mathbf{h}^{n}$ we have $P_e^{(n)}\rightarrow 0$ as $n\rightarrow\infty$. The capacity, $\Cc$, is the supremum of all achievable rates $R$.
\end{defi}

\textbf{Memoryless active adversary:} The mapping at each adversary satisfies:
\begin{align}
p_j(\mathbf{x}_a^n(j)|\mathbf{x}^n(j),\mathbf{s}^n(j))=\prod\limits_{i=1}^{n}p_j(\mathbf{x}_{a,i}(j)|\mathbf{x}_i(j),\mathbf{s}_i(j))\label{eqn:pmf_iid}
\end{align}
i.e., the adversary uses the same probability distribution to modify the transmitted symbols in each channel use. For each route  $j$, the distribution in \eqref{eqn:pmf_iid} is independent and identically distributed (i.i.d) and fixed over time; but, clearly, the distributions can differ across routes.

\textbf{Foreseer active adversary:} It observes the transmitted codeword over the entire block (i.e., $\mathbf{x}^n(j)$) upon which it bases its strategy. That is, while satisfying \eqref{eqn:pmf3}, the adversary can choose the position and value of the symbols in the codeword to be modified. In this case, we concentrate on two types of attacks:

\emph{Replacement attacks}: $\Xc=\Xc_a=\Yc$ with hamming distortion measure:
\begin{equation}\label{eqn:ham_dis}
    \setlength{\nulldelimiterspace}{0pt}
    d(x,\hat{x})=\left\{\begin{IEEEeqnarraybox}[\relax][c]{l's}
    1,&if $x\neq\hat{x}$\\
    0,&if $x=\hat{x}$%
    \end{IEEEeqnarraybox}\right.
\end{equation}

\emph{Erasing (dropping) attacks} (also known as selective forwarding): $\Xc_a,\Yc=\{\Xc,e\}$ where for all $x,x'\in\Xc,x\neq x'$, $d(x,x)=0$, $d(x,x')=\infty$ and $d(e,x)=d(x,e)=1$. With this definition, we limit the adversaries only to erase the data and they cannot replace data as long as their distortion limits are finite, i.e., $D_j<\infty$ for $j\in\{1,\ldots,n_r\}$.

These two types cover all possible modification attacks. It is reasonable to assume that anything outside the alphabet is rejected by Rx; thus, this can be modeled as an erased symbol. Therefore, the adversary does not gain anything by modifying to a non-existent symbol.

\section{Main Results}\label{sec:main}
For our multi-route PP-MA, for both memoryless and foreseer adversaries, we consider either no CSI or CSI at Tx. The adversaries are assumed to have perfect CSI.

\subsection{Memoryless active adversaries}\label{subsec:main_iid}
We state the capacity for the channel in \eqref{eqn:pmf_iid}, first assuming no CSI available at the Tx and Rx.
\begin{theorem}\label{thm:cap_iid_noCSI}
The capacity of the multi-route PP-MA satisfying \eqref{eqn:pmf_iid}, with no CSI available at either the Tx or the Rx is:
\begin{IEEEeqnarray}{l}
\Cc_{i.i.d}^{\text{nC}}=\sup_{p(\mathbf{x})}\;\min_{\substack{\mathbf{s}\in\{0,1\}^{n_r}\\w_H(\mathbf{s})\leq n_a}}
\inf_{\substack{\prod\limits_{j=1}^{n_r}p_{j}(\mathbf{x}_{a}(j)|\mathbf{x}(j),\mathbf{s}(j))\\\forall j\in\{1,\ldots,n_r\}:E_{p_j}[d(\mathbf{X}_{a}(j),\mathbf{X}(j))]\leq D_{j,s}=s\cdot D_j}}
\sum\limits_{j=1}^{n_r}I(\mathbf{X}(j);\mathbf{Y}_s(j))\label{eqn:cap_iid_noCSI}
\end{IEEEeqnarray}
where
$\forall\mathbf{s}\in\{0,1\}^{n_r}$, we have $\mathbf{Y}_s\in\Yc^{n_r}$ and
\begin{IEEEeqnarray*}{l}
p(\mathbf{y}^n|\mathbf{x}^n,\mathbf{x}_a^n,\mathbf{s}^n)=\prod\limits_{j=1}^{n_r}p_{j\,Y|X_a,X,S}(\mathbf{y}^n(j)|\mathbf{x}^n(j),\mathbf{x}_a^n(j),\mathbf{s}^n(j))=\prod\limits_{j=1}^{n_r}
p_{j\,Y_s|X_a,X}(\mathbf{y}^n(j)|\mathbf{x}^n(j),\mathbf{x}_a^n(j)).
\end{IEEEeqnarray*}
Hence, the mutual information term is evaluated with respect to the joint p.m.f \eqref{eqn:pmf}.
\end{theorem}
\begin{IEEEproof}[Proof outline]
For the achievablility part, we use a random coding argument in a compound channel (to model the adversaries' placement). To take into account all possible i.i.d adversaries' strategies, we consider all possible joint types of $(\mathbf{x}^{n}(j),\mathbf{y}^{n}(j))$ for the $j$-th route, subject to the distortion limit on $\mathbf{x}_a^{n}(j)$.
The converse follows from Fano's inequality, by noting that for \emph{every} adversaries' placement and mapping ($\mathbf{s}$ and $\mathbf{h}^{n}$) we must have $P_e^{(n)}\stackrel{n\rightarrow\infty}{\longrightarrow} 0$. Detailed proof in Appendix.
\end{IEEEproof}
\begin{remark}
On the $j$-th route, the conditional distribution of the adversary's channel input, i.e., $p_{j}(\mathbf{x}_{a}(j)$ $|\mathbf{x}(j),\mathbf{s}(j))$, can model all possible memoryless active attacks (e.g., replacement or dropping). To specify a certain attack, it is enough to properly define the input alphabets, $\Xc_a,\Xc$, and the distortion measure $d(.,.)$. Thus, the $\inf$ is calculated over a feasible set of $\Xc\times\Xc_a$ distributions ($p_j$), where the feasibility constraint is determined by $E_{p_j}[d(\mathbf{X}_{a}(j),\mathbf{X}(j))]\leq D_{j,s}=s\cdot D_j$.
\end{remark}

Next, we obtain the capacity when CSI is available at Tx (proof in Appendix).
\begin{theorem}\label{thm:cap_iid_TxCSI}
The capacity of multi-route PP-MA satisfying \eqref{eqn:pmf_iid}, with CSI available at Tx is:
\begin{IEEEeqnarray}{l}
\Cc_{i.i.d}^{\text{TC}}=\min_{\substack{\mathbf{s}\in\{0,1\}^{n_r}\\w_H(\mathbf{s})\leq n_a}}\;\sup_{p(\mathbf{x})}\;
\inf_{\substack{\prod\limits_{j=1}^{n_r}p_{j}(\mathbf{x}_{a}(j)|\mathbf{x}(j),\mathbf{s}(j))\\\forall j\in\{1,\ldots,n_r\}:E_{p_j}[d(\mathbf{X}_{a}(j),\mathbf{X}(j))]\leq D_{j,s}=s\cdot D_j}}
\sum\limits_{j=1}^{n_r}I(\mathbf{X}(j);\mathbf{Y}_s(j))\label{eqn:cap_iid_TxCSI}
\end{IEEEeqnarray}
where the notation $Y_s$ is defined in Theorem~\ref{thm:cap_iid_noCSI} and the mutual information term is evaluated with respect to the joint p.m.f \eqref{eqn:pmf}.
\end{theorem}

\subsection{Foreseer active adversaries}\label{subsec:main_niid}
Now, we derive lower and upper bounds on the capacity for all possible foreseer adversaries strategies. The bounds are based on the possible minimum distances the legitimate user codewords can tolerate under each attack.
\begin{theorem}\label{thm:low_mem_noCSI}
A lower bound to the capacity of the multi-route PP-MA with foreseer adversaries (no CSI available at Tx or Rx) is:
\begin{IEEEeqnarray*}{l}
\Rc_{l}^{\text{nC}}=\sup\min\inf_{\mathbf{h}^n}\sum\limits_{j=1}^{n_r}[H(\mathbf{V})-H(\mathbf{X}_a(j)|\mathbf{Y}(j))-\frac{H_{|\Xc|}(d_j)}{\log_{|\Xc|} 2} ]^+
\end{IEEEeqnarray*}
where the supremum and the minimum are taken over $p(\mathbf{x})p(\mathbf{v}|\mathbf{x});\forall j\in\{1,\ldots,n_r\}:E_{p_j}[d(\mathbf{V}(j),\mathbf{X}(j))]\leq d_{j}$ and $\mathbf{s}\in\{0,1\}^{n_r}:w_H(\mathbf{s})\leq n_a$, respectively; $d_j=f(D_{j,s}=\mathbf{s}(j)\cdot D_j)$ is determined based on the attack type and the distortion measure (i.e., $d_j=\mathbf{s}(j)\cdot2D_j$ for replacement attacks and $d_j=\mathbf{s}(j)\cdot D_j$ for erasing attacks); the second entropy, $H(\mathbf{X}_a(j)|\mathbf{Y}(j))$, is evaluated with respect to the memoryless channel: $p_j(\mathbf{y}^n(j)|\mathbf{x}_a^n(j))=\prod\limits_{i=1}^{n}p_j(\mathbf{y}_i(j)|\mathbf{x}_{a,i}(j))$.
\end{theorem}
\begin{IEEEproof}[Proof outline] We apply a random coding technique on top of a random linear code (Varshamov construction \cite{Var57}), by introducing proper auxiliary codewords. Random coding is used to combat the stochastic behavior of the $\Xc_a\mapsto\Yc$ channel. Varshamov construction guarantees recovery from the worst-case errors, by making the minimum distance of the code greater than the number of errors. First, we generate auxiliary codewords, $\mathbf{u}$; then, we apply a random linear coding $n_r$ times to these codewords, to generate the transmitted codewords, $\mathbf{x}^{n}$. To decode from the $j$-th route: if Rx can decode the adversary's channel input $\mathbf{x}_a^{n}(j)$, the transmitted codeword is the only $\mathbf{x}^{n}(j)$ in a Hamming ball with radius $d_j$. To apply this scheme, we choose $\mathbf{v}^{n}(j)$ as the possible $\mathbf{x}_a^{n}(j)$ and try to decode it after receiving $\mathbf{y}^n(j)$, by decreasing its rate to satisfy the stochastic limitation imposed by the $\Xc_a\mapsto\Yc$ channel. Proof details in the Appendix.
\end{IEEEproof}
\begin{theorem}\label{thm:low_mem_TxCSI}
A lower bound on the capacity of the multi-route PP-MA with foreseer adversaries (CSI available at Tx) is:
\begin{IEEEeqnarray*}{l}
\Rc_{l}^{\text{TC}}=\min\sup\inf_{\mathbf{h}^n}\sum\limits_{j=1}^{n_r}[H(\mathbf{V})-H(\mathbf{X}_a(j)|\mathbf{Y}(j))-\frac{H_{|\Xc|}(d_j)}{\log_{|\Xc|} 2}]^+
\end{IEEEeqnarray*}
where the minimum and the supremum are taken over $\mathbf{s}\in\{0,1\}^{n_r}:w_H(\mathbf{s})\leq n_a$ and $p(\mathbf{x})p(\mathbf{v}|\mathbf{x});\forall j\in\{1,\ldots,n_r\}:E_{p_j}[d(\mathbf{V}(j),\mathbf{X}(j))]\leq d_{j}$, respectively; $d_j$ and $H_q(x)$ are defined in Theorem~\ref{thm:low_mem_noCSI}; the second entropy, $H(\mathbf{X}_a(j)|\mathbf{Y}(j))$, is evaluated with respect to $p_j(\mathbf{y}^n(j)|\mathbf{x}_a^n(j))=\prod\limits_{i=1}^{n}p_j(\mathbf{y}_i(j)|\mathbf{x}_{a,i}(j))$.
\end{theorem}

\begin{remark}
In both $\Rc_{l}^{\text{nC}}$ and $\Rc_{l}^{\text{TC}}$, the first term is independent of $\mathbf{s}$ and the second term is independent of $p(\mathbf{x})$. Therefore, we have $\Rc_{l}^{\text{nC}}=\Rc_{l}^{\text{TC}}$. That is, CSI does not help the achieving strategy for these rates.
\end{remark}

\begin{theorem}\label{thm:up_mem}
The following are upper bounds to the capacity of the multi-route PP-MA with foreseer adversaries:
\begin{IEEEeqnarray}{rcl}
\Rc_{u}^{\text{nC}}&=&\sup_{p(\mathbf{x})}\min\inf_{\mathbf{h}^n}\;\sum\limits_{j=1}^{n_r}[I(\mathbf{X}_a(j);\mathbf{Y}(j))-\frac{H_{|\Xc|}(\frac{d_j}{2})}{\log_{|\Xc|} 2}]^+\quad\;\;\label{eqn:up_mem_noCSI}\\
\Rc_{u}^{\text{TC}}&=&\min\sup_{p(\mathbf{x})}\inf_{\mathbf{h}^n}\;\sum\limits_{j=1}^{n_r}[I(\mathbf{X}_a(j);\mathbf{Y}(j))-\frac{H_{|\Xc|}(\frac{d_j}{2})}{\log_{|\Xc|} 2}]^+\label{eqn:up_mem_TxCSI}
\end{IEEEeqnarray}
where the minimum is taken over $\mathbf{s}\in\{0,1\}^{n_r}:w_H(\mathbf{s})\leq n_a$ and $d_j$ and $H_q(x)$ are defined in Theorem~\ref{thm:low_mem_noCSI}.
\end{theorem}
\begin{IEEEproof}[Proof outline]
We follow an approach similar to the one used to derive the Hamming bound, that is, we limit the volume of the coding balls. Proof in the Appendix.
\end{IEEEproof}

\section{Examples}\label{sec:examples}

\textbf{Replacement attacks to binary transmission:} The channel inputs and output have binary alphabets (i.e., $\Xc,\Xc_a,\Yc=\{0,1\}$) and $d$ is the Hamming distortion measure (defined in Section~\ref{sec:model}). The stochastic channel from the adversary to the Rx is assumed to be a Binary Symmetric Channel (BSC). Thus, the channel output at Rx at time $i\in\{1,\ldots,n\}$ is:
\begin{IEEEeqnarray}{rcl}
\mathbf{Y}_i(j)&\:=\:&\mathbf{X}_{a,i}(j)\oplus \mathbf{Z}_i(j)\label{eqn:BSC_ch}
\end{IEEEeqnarray}
where $\mathbf{Z}_i(j)\sim \Bc(N_j)$ for $j\in\{1,\ldots,n_r\}$.
First, consider memoryless active adversaries. We obtain the results of Theorems~\ref{thm:cap_iid_noCSI} and \ref{thm:cap_iid_TxCSI} as:
\begin{corollary}\label{corr:cap_iid_BSC}
The capacity of the multi-route PP-MA with binary alphabets satisfying \eqref{eqn:pmf_iid}, \eqref{eqn:BSC_ch}, for both no CSI and CSI at Tx, is:
\begin{IEEEeqnarray*}{l}
\Cc_{i.i.d}^{\text{nC}}=\Cc_{i.i.d}^{\text{TC}}=n_r-\max_{\substack{\mathbf{s}\in\{0,1\}^{n_r}\\w_H(\mathbf{s})\leq n_a}}\sum\limits_{j=1}^{n_r}
\max_{N'_j\leq \mathbf{s}(j)\cdot \tilde{D}_j}H(N_j\ast N'_j)
\end{IEEEeqnarray*}
where $\tilde{D}_j=\min\{D_j,1-D_j\}$ and $\alpha\ast\beta=\alpha(1-\beta)+\beta(1-\alpha)$. If we assume identical route conditions, $D_j=D\leq\frac{1}{2}$ and $N_j=N$ for $j\in\{1,\ldots,n_r\}$, the capacity is: $n_r-(n_r-n_a)H(N)-n_aH(N\ast D)$.
\end{corollary}

\begin{IEEEproof}
Let $P_j=Pr(\mathbf{X}(j)=1)$ and without loss of generality assume $P_j\leq \frac{1}{2}$. To find the $\inf\sum\limits_{j=1}^{n_r}I(\mathbf{X}(j);\mathbf{Y}_s(j))$ in \eqref{eqn:cap_iid_noCSI}, we first find a lower bound to it and we then show it is achievable by the adversaries.
\begin{IEEEeqnarray*}{rcl}
I(\mathbf{X}(j);\mathbf{Y}_s(j))&=&H(\mathbf{X}(j))-H(\mathbf{X}(j)|\mathbf{Y}_s(j))\\
&{\geq}& H(P_j)-H(\mathbf{X}(j)\oplus\mathbf{X}_a(j)\oplus\mathbf{Z}(j))\stackrel{(a)}{\geq}H(P_j)-H(N_j\ast N'_j)
\end{IEEEeqnarray*}
in (a) we define $N'_j\leq \mathbf{s}(j)\cdot D_j$ and use $Pr(\mathbf{X}(j)\neq\mathbf{X}_a(j))\leq \mathbf{s}(j)\cdot D_j$. This lower bound is achievable by the $j$-th adversary if it chooses a joint distribution given by two backward BSCs, $\Yc\rightarrow\Xc_a$ and $\Xc_a\rightarrow\Xc$, with cross-over probabilities $N_j$ and $N'_j$, respectively. This results in $Pr(\mathbf{X}_a(j)=1)=\frac{P_j-N'_j}{1-2N'_j}$. Hence, we need $P_j\geq N'_j$ to hold. Therefore, \eqref{eqn:cap_iid_noCSI} for this channel is:
\begin{IEEEeqnarray*}{l}
\Cc_{i.i.d}^{\text{nC}}=
\sup_{0 \leq D_j \leq P_j\leq \frac{1}{2}}\min_{\substack{\mathbf{s}\in\{0,1\}^{n_r}\\w_H(\mathbf{s})\leq n_a}}\;\sum\limits_{j=1}^{n_r}[H(P_j)-
\max_{N'_j\leq \mathbf{s}(j)\cdot D_j}H(N_j\ast N'_j)]
\end{IEEEeqnarray*}
The rest of the proof is straightforward.
\end{IEEEproof}

Now, consider foreseer active adversaries. We obtain the results of Theorems~\ref{thm:low_mem_noCSI} and \ref{thm:low_mem_TxCSI} as:
\begin{corollary}\label{corr:low_mem_BSC}
The lower bound to the capacity of the multi-route PP-MA with foreseer adversaries, binary alphabets satisfying \eqref{eqn:BSC_ch}, for both no CSI and CSI at Tx is:
\begin{IEEEeqnarray}{l}
\Rc_{l}^{\text{nC}}=\Rc_{l}^{\text{TC}}=
n_r-\sum\limits_{j=1}^{n_r}H(N_j)-\max_{\substack{\mathbf{s}\in\{0,1\}^{n_r}\\w_H(\mathbf{s})\leq n_a}}\sum\limits_{j=1}^{n_r}H(\mathbf{s}(j)\cdot 2D_j)\label{eqn:BSC_low_mem}
\end{IEEEeqnarray}
For identical route conditions, $D_j=D\leq\frac{1}{2}$ and $N_j=N$ for $j\in\{1,\ldots,n_r\}$, the rate is $n_r(1-H(N))-n_aH(2D)$.
\end{corollary}

\begin{IEEEproof}
Let $\mathbf{V}_i(j)=\mathbf{X}_{i}(j)$ and $P_j=Pr(\mathbf{X}(j)=1)$.
Recall that for all $\mathbf{h}^n$ (Definition~\ref{def:code}, satisfying \eqref{eqn:pmf3}), we have $Pr(\mathbf{X}(j)\neq\mathbf{X}_a(j))\leq \mathbf{s}(j)\cdot D_j$. After some calculations, we can compute $H(\mathbf{V}(j))=H(P_j)$ and
\begin{IEEEeqnarray*}{rcl}
H(\mathbf{X}_a(j)|\mathbf{Y}(j))&\leq& H(\mathbf{X}_a(j)\oplus \mathbf{Y}(j))=H(\mathbf{Z}_i(j))=H(N_j)
\end{IEEEeqnarray*}
and obtain \eqref{thm:low_mem_noCSI} as:
\begin{IEEEeqnarray}{rcl}
\Rc_{l}^{\text{nC}}&\geq&\sup_{0 \leq P_j\leq \frac{1}{2}}\min_{\substack{\mathbf{s}\in\{0,1\}^{n_r}\\w_H(\mathbf{s})\leq n_a}}\sum\limits_{j=1}^{n_r}[H(P_j)-H(N_j)-H(\mathbf{s}(j)\cdot 2D_j)]\label{eqn:BSC_low_mem_p1}
\end{IEEEeqnarray}
which will be maximized for $P_j=\frac{1}{2}$ independently of $\mathbf{s}(j)$, for $j\in\{1,\ldots,n_r\}$. This results in \eqref{eqn:BSC_low_mem}. It is easy to see that computing $\Rc_{l}^{\text{TC}}$ in Theorem~\ref{thm:low_mem_TxCSI} results in the same rate.
\end{IEEEproof}

We adapt Theorem~\ref{thm:up_mem} for binary alphabets and the BSC of \eqref{eqn:BSC_low_mem}:
\begin{corollary}\label{corr:up_mem_BSC}
The upper bound to the capacity of the multi-route PP-MA with foreseer adversaries, binary alphabets satisfying \eqref{eqn:BSC_ch}, for both no CSI and CSI at Tx, is:
\begin{IEEEeqnarray}{l}
\Rc_{u}^{\text{nC}}=\Rc_{u}^{\text{TC}}=
n_r-\sum\limits_{j=1}^{n_r}H(N_j)-\max_{\substack{\mathbf{s}\in\{0,1\}^{n_r}\\w_H(\mathbf{s})\leq n_a}}\sum\limits_{j=1}^{n_r}H(\mathbf{s}(j)\cdot D_j)\label{eqn:BSC_up_mem}
\end{IEEEeqnarray}
For identical route conditions, $D_j=D$ and $N_j=N$ for $j\in\{1,\ldots,n_r\}$, the rate is $n_r(1-H(N))-n_aH(D)$.
\end{corollary}
\begin{IEEEproof}
We combine the methods of Corollaries~\ref{corr:cap_iid_BSC} and \ref{corr:low_mem_BSC}. We can show that the sum of the first and the second terms in the right side of \eqref{eqn:BSC_low_mem_p1} makes an upper bound on the first term of \eqref{eqn:up_mem_noCSI} and \eqref{eqn:up_mem_TxCSI}. To do this, it is enough to choose the proper joint distribution for the adversaries' input that achieves this bound. This distribution consists of two backward BSCs, $\Yc\rightarrow\Xc_a$ and $\Xc_a\rightarrow\Xc$, with cross-over probabilities $N_j$ and $D_j$, respectively. The rest of the proof is similar to that of Corollary~\ref{corr:low_mem_BSC}.
\end{IEEEproof}

\textbf{Erasing attacks on binary transmission:} To reduce the erasing attacks to binary alphabets, we set: $\Xc=\{0,1\}$, $\Xc_a,\Yc=\{0,1,e\}$, $d(0,0)=d(1,1)=0$, $d(0,1)=d(1,0)=\infty$, and $d(0,e)=d(1,e)=1$.
Across the $\Xc_a\mapsto\Yc$ channel, additional erasing is introduced for the received signal at Rx (not distinguishable from the adversarial erasing at Rx).
Thus, the channel output at Rx at time $i\in\{1,\ldots,n\}$ is:
\begin{equation}\label{eqn:BEC_ch}
\setlength{\nulldelimiterspace}{0pt}
\mathbf{Y}_i(j)=\left\{\begin{IEEEeqnarraybox}[\relax][c]{l's}
\text{BEC}(\mathbf{X}_{a,i}(j), N_j),&if $\mathbf{X}_{a,i}(j)\neq e$\\
\mathbf{X}_{a,i}(j),&if $\mathbf{X}_{a,i}(j)= e$%
\end{IEEEeqnarraybox}\right.
\end{equation}
where BEC$(x,\beta)$ shows a Binary Erasure Channel (BEC) with input $x$ and probability of erasure $\beta$.
Here, we state our results for both memoryless and foreseer adversaries. Proofs in Appendix.
\begin{corollary}\label{corr:cap_iid_BEC}
The capacity of the multi-route PP-MA with $\Xc=\{0,1\}$ and $\Xc_a,\Yc=\{0,1,e\}$, satisfying \eqref{eqn:pmf_iid} and \eqref{eqn:BEC_ch}, for both no CSI and CSI at Tx, is:
\begin{IEEEeqnarray*}{l}
\Cc_{i.i.d}^{\text{nC}}=\Cc_{i.i.d}^{\text{TC}}=\min_{\substack{\mathbf{s}\in\{0,1\}^{n_r}\\w_H(\mathbf{s})\leq n_a}}\sum\limits_{j=1}^{n_r}(1-\mathbf{s}(j)\cdot D_j)(1-N_j).
\end{IEEEeqnarray*}
For identical route conditions, $D_j=D$ and $N_j=N$ for $j\in\{1,\ldots,n_r\}$, the capacity is $(1-N)(n_r-n_aD)$.
\end{corollary}
\begin{corollary}\label{corr:low_mem_BEC}
The lower bound to the capacity of the multi-route PP-MA with foreseer adversaries, $\Xc=\{0,1\}$ and $\Xc_a,\Yc=\{0,1,e\}$, satisfying \eqref{eqn:BEC_ch}, for both no CSI and CSI at Tx, is:
\begin{IEEEeqnarray*}{l}
\Rc_{l}^{\text{nC}}=\Rc_{l}^{\text{TC}}=
\min_{\substack{\mathbf{s}\in\{0,1\}^{n_r}\\w_H(\mathbf{s})\leq n_a}}\sum\limits_{j=1}^{n_r}1-N_j(1-N'_j)-\bar{N}_jH(\frac{N'_j}{\bar{N}_j})-H(\mathbf{s}(j)\cdot D_j)
\end{IEEEeqnarray*}
where $\bar{N}_j=N_j(1-\mathbf{s}(j)\cdot D_j)+\mathbf{s}(j)\cdot D_j$.
For identical route conditions, $D_j=D$ and $N_j=N$ for $j\in\{1,\ldots,n_r\}$, the rate is $n_r(1-N)-n_a[(N(1-D)+D)H(\frac{D}{N(1-D)+D})+H(D)-D]$.
\end{corollary}
\begin{corollary}\label{corr:up_mem_BEC}
The upper bound to the capacity of the multi-route PP-MA with foreseer adversaries, $\Xc=\{0,1\}$ and $\Xc_a,\Yc=\{0,1,e\}$, satisfying \eqref{eqn:BEC_ch}, for both no CSI and CSI at Tx, is:
\begin{IEEEeqnarray}{l}
\Rc_{u}^{\text{nC}}=\Rc_{u}^{\text{TC}}=
\min_{\substack{\mathbf{s}\in\{0,1\}^{n_r}\\w_H(\mathbf{s})\leq n_a}}\sum\limits_{j=1}^{n_r}[H((1-N'_j)(1-N_j))+(1-N'_j)(1-N_j-H(N_j))-H(\mathbf{s}(j)\cdot \frac{D_j}{2})]\nonumber
\end{IEEEeqnarray}
where $N'_j=\mathbf{s}(j)\cdot D_j$. For identical route conditions, $D_j=D$ and $N_j=N$ for $j\in\{1,\ldots,n_r\}$, the rate is $n_a(H((1-N)(1-D))+(1-D)(1-N-H(N))-H(\frac{D}{2}))+(n_r-n_a)(1-N)$.
\end{corollary}

\textbf{Gaussian replacement attacks:} We assume Gaussian distributions for the channel inputs and output. The distortion measure now is the squared error distortion:
\begin{equation*}
d(x,\hat{x})= (x-\hat{x})^2
\end{equation*}
and the channel model can be shown as:
\begin{IEEEeqnarray}{rcl}
\mathbf{Y}_i(j)&\:=\:&\mathbf{X}_{a,i}(j)+\mathbf{Z}_i(j)\label{eqn:Gaus_ch}
\end{IEEEeqnarray}
where $\mathbf{Z}_i(j)\sim \Nc(0,N_j)$ for $j\in\{1,\ldots,n_r\}$ are independent and i.i.d Gaussian noise components. We assume the average power constraint on input signal  $\mathbf{X}(j)$ as $\frac{1}{n}\sum\limits_{t=1}^n|\mathbf{x}_{t}(j)|^2\leq \mathbf{P}_j$. Hence, $\mathbf{X}(j)\sim \Nc(0,P_j)$ for $j\in\{1,\ldots,n_r\}$.
Here, we only consider the memoryless adversaries and obtain the results of Theorems~\ref{thm:cap_iid_noCSI} and \ref{thm:cap_iid_TxCSI} (proof in Appendix).
\begin{corollary}\label{corr:cap_iid_Gaus}
The capacity of the multi-route PP-MA with Gaussian distributions for channel inputs and output, satisfying \eqref{eqn:pmf_iid} and \eqref{eqn:Gaus_ch}, for both no CSI and CSI at Tx, is:
\begin{IEEEeqnarray*}{l}
\Cc_{i.i.d}^{\text{nC}}=\Cc_{i.i.d}^{\text{TC}}=\max_{\substack{\mathbf{s}\in\{0,1\}^{n_r}\\w_H(\mathbf{s})\leq n_a}}\sum\limits_{j=1}^{n_r}
\theta(\frac{P_j-\mathbf{s}(j)\cdot D_j+N_j}{\mathbf{s}(j)\cdot D_j+N_j})
\end{IEEEeqnarray*}
where $\theta(x)\doteq \frac{1}{2}\log(x)$. For identical route conditions, $D_j=D$ and $N_j=N$ with equal power constraints $P_j=P$ for $j\in\{1,\ldots,n_r\}$, the capacity is: $n_r\theta(1+\frac{P}{N})-n_a\theta(1+\frac{D(P+2N)}{N(P-D+N)})$.
\end{corollary}

\begin{table*}
\renewcommand{\arraystretch}{1.3}
\caption{Our results for the replacement and erasing attacks on binary transmission with $n_r=n_a=1$.}
\label{tbl:comparison} \centering
\begin{tabular}{|c|c|c|c|}
\hline
&& Replacement   &  Erasing\\
\hline
Memoryless&Capacity& $1-H(N\ast D)$  & $(1-N)(1-D)$\\
\hline
Foreseer &lower& $1-H(N)-H(2D)$ & $1-N(1-D)-(N(1-D)+D)H(\frac{D}{N(1-D)+D})-H(D)$\\
\cline{2-4}
&upper& $1-H(N)-H(D)$ & $H((1-N)(1-D))+(1-D)(1-N-H(N))-H(\frac{D}{2})$\\
\hline
\end{tabular}
\end{table*}

\textbf{Comparison:} Along with identical route conditions, to simplify, let $n_r=n_a=1$. Table~\ref{tbl:comparison} shows the results for the replacement and erasing attacks on binary transmission. Obviously, for zero distortion for the adversary ($D=0$), we have BSC and BEC with parameter $N$. The rate reduction caused by a foreseer adversary is considerable. Consider only the adversary's effect by setting $N=0$: the foreseer is twice more powerful than the memoryless one (in terms of the lower bound) for the replacement attack. For the erasing attack, the foreseer reduces (compared to the memoryless) the rate from a BEC rate (i.e., $1-D$) to a BSC rate (i.e., $1-H(D)$).
For Gaussian replacement attacks (under these simplified assumptions), the capacity is $\frac{1}{2}\log(1+\frac{P-2D}{D+N})$; while, for Gaussian independent jamming with power $D$, we achieve $\frac{1}{2}\log(1+\frac{P}{D+N})$. Thus, knowing the transmitted codeword (even in a memoryless case) worsens the situation compared to an independent jammer.

\bibliographystyle{./IEEEtran}
\bibliography{./IEEEabrv,./ISIT2014}

\appendix

\begin{IEEEproof}[Proof of Theorem~\ref{thm:cap_iid_noCSI}]

\textit{Achievability:}
We use random encoding and joint typicality decoding. Considering the problem setup in Section~\ref{sec:model}, we denote the set of possible joint types of triple of sequences $(x^n,x^n_{a},y^n)\in\Xc^n\times\Xc^n_a\times\Yc^n$ as:
\begin{IEEEeqnarray*}{l}
\Pc^n_{j,s}(\Xc\times\Xc_a\times\Yc)=\{\pi_j(x,x_a,y|x^n,x_{a}^n,y^n): E_{p_{j,s}^n}[d(X_{a}^n,X^n)]\leq D_{j,s}=s\cdot D_j\}\yesnumber\label{eqn:type3}
\end{IEEEeqnarray*}
and the possible pairs of $(x^n,y^n)\in\Xc^n\times\Yc^n$ for some $x^n_{a}\in\Xc^n_a$:
\begin{IEEEeqnarray*}{l}
\Pc^n_{j,s}(\Xc\times\Yc)=\{\pi_j(x,y|x^n,y^n):\exists x_{a}^n\in\Xc^n_a \text{ such that } (x^n,x_{a}^n,y^n)\in\Pc^n_{j,s}(\Xc\times\Xc_a\times\Yc)\}\yesnumber\label{eqn:type2}
\end{IEEEeqnarray*}
For $q\in\Pc^n_{j,s}(\Xc\times\Yc)$, the type class is defined as $\Tc^n_{j,s}(q)=\{(x^n,y^n), p_{XY}(x,y)=q\}$.
Note that, given the adversaries are memoryless, we have:
\begin{IEEEeqnarray*}{rcl}
p(\mathbf{y}_s^n|\mathbf{x}^n)&=&\prod\limits_{i=1}^{n}p_{\mathbf{Y}_{s}|\mathbf{X}}(\mathbf{y}_{s,i}|\mathbf{x}_i)=\prod\limits_{i=1}^{n}p_{\mathbf{Y}|\mathbf{X},\mathbf{S}}(\mathbf{y}_{s,i}|\mathbf{x}_i,\mathbf{s})\\
&=&\sum\limits_{\mathbf{x}_{a}}\prod\limits_{i=1}^{n}\prod\limits_{j=1}^{n_r}q_{j,s}(\mathbf{x}_{a,i}(j)|\mathbf{x}_i(j))p_j(\mathbf{y}_i(j)|\mathbf{x}_{a,i}(j))\yesnumber\label{eqn:pmf_iid_noCSI}
\end{IEEEeqnarray*}
where $q_{j,s}$ is defined in \eqref{eqn:pmf3}.

Fix $p_{\mathbf{X}}(\mathbf{x})$ and generate $2^{nR}$ i.i.d sequences $\mathbf{x}^{n}[m]$, each with probability $\prod\limits_{i=1}^np_{\mathbf{X}}(\mathbf{x}_{i})$, where $m\in[1:2^{nR}]$. To transmit $m$, Tx sends $\mathbf{x}^{n}[m]$. Rx after receiving $\mathbf{y}^n$, looks for a unique index $\tilde{m}$ that satisfies:
\begin{IEEEeqnarray*}{c}
(\mathbf{x}^{n}[\tilde{m}],\mathbf{y}^n)\in A_\epsilon^{n}(\mathbf{X},\mathbf{Y}_s).
\end{IEEEeqnarray*}

Due to the symmetry of the random codebook generation, the probability of error is independent of the specific messages. Hence, to analyze the probability of error, without loss of generality, we assume that $m=1$ is encoded and transmitted. The error events at Rx are:
\begin{IEEEeqnarray*}{rcl}
\Ec_1&=&\{\forall\mathbf{s}\in\{0,1\}^{n_r}, w_H(\mathbf{s})\leq n_a: (\mathbf{x}^{n}[1],\mathbf{y}^n)\notin A_\epsilon^{n}(\mathbf{X},\mathbf{Y}_s)\}\\
\Ec_2&=&\{\exists\mathbf{s}\in\{0,1\}^{n_r}, w_H(\mathbf{s})\leq n_a \text{ such that } (\mathbf{x}^{n}[m],\mathbf{y}^n)\in A_\epsilon^{n}(\mathbf{X},\mathbf{Y}_s) \text{ for some } m\neq 1\}
\label{eqn:ErrEve_iid_noCSI}
\end{IEEEeqnarray*}

Due to the Asymptotic Equipartition Property (AEP) \cite{CovTho06}, $Pr\left(\Ec_1\right)\rightarrow 0$ as $n\rightarrow\infty$. Now, to consider the probability of $\Ec_2$, let $\forall\mathbf{s}\in\{0,1\}^{n_r}$:
\begin{IEEEeqnarray*}{rcl}
\Ec'_{2,\mathbf{s}}&=&\{\exists m\neq 1 \text{ such that } (\mathbf{x}^{n}[m],\mathbf{y}^n)\in A_\epsilon^{n}(\mathbf{X},\mathbf{Y}_s)\}
\yesnumber\label{eqn:ErrEve2_iid_noCSI}
\end{IEEEeqnarray*}
with probability:
\begin{IEEEeqnarray*}{rcl}
Pr(\Ec'_{2,\mathbf{s}})&\stackrel{(a)}{\leq}&\sum_{m\neq 1}\prod\limits_{j=1}^{n_r}\sum_{q\in\Pc^n_{j,s}(\Xc\times\Yc)}Pr((\mathbf{x}^{n}(j)[m],\mathbf{y}^{n}(j))\in\Tc^n_{j,s}(q))\\
&\leq&\sum_{m\neq 1}\prod\limits_{j=1}^{n_r}|\Pc^n_{j,s}(\Xc\times\Yc)|\sup_{q\in\Pc^n_{j,s}(\Xc\times\Yc)}Pr((\mathbf{x}^{n}(j)[m],\mathbf{y}^{n}(j))\in\Tc^n_{j,s}(q))\\
&\stackrel{(b)}{\leq}&\sum_{m\neq 1}\prod\limits_{j=1}^{n_r}|\Pc^n_{j,s}(\Xc\times\Yc)|\sup_{\substack{p_{j}(\mathbf{x}_{a}(j)|\mathbf{x}(j),\mathbf{s}(j))\\E_{p_j}[d(\mathbf{X}_{a}(j),\mathbf{X}(j))]\leq D_{j,s}=s\cdot D_j}}2^{-n(I(\mathbf{X}(j);\mathbf{Y}_s(j))-\delta(\epsilon))}\\
&\stackrel{(c)}{\leq}&\sum_{m\neq 1}\prod\limits_{j=1}^{n_r}\left(\!\!\binom{|\Xc|}{n}\!\!\right)\left(\!\!\binom{|\Yc|}{n}\!\!\right)\left(\!\!\binom{|\Xc_a|}{n}\!\!\right)\sup_{\substack{p_{j}(\mathbf{x}_{a}(j)|\mathbf{x}(j),\mathbf{s}(j))\\E_{p_j}[d(\mathbf{X}_{a}(j),\mathbf{X}(j))]\leq D_{j,s}=s\cdot D_j}}2^{-n(I(\mathbf{X}(j);\mathbf{Y}_s(j))-\delta(\epsilon))}\\
&{\leq}&\sum_{m\neq 1}\prod\limits_{j=1}^{n_r}\frac{n^{|\Xc|+|\Yc|+|\Xc_a|}}{(|\Xc|-1)!(|\Yc|-1)!(|\Xc_a|-1)!}\sup_{\substack{p_{j}(\mathbf{x}_{a}(j)|\mathbf{x}(j),\mathbf{s}(j))\\E_{p_j}[d(\mathbf{X}_{a}(j),\mathbf{X}(j))]\leq D_{j,s}=s\cdot D_j}}2^{-n(I(\mathbf{X}(j);\mathbf{Y}_s(j))-\delta(\epsilon))}\\
&\stackrel{(d)}{\leq}& n^{|\Xc|+|\Yc|+|\Xc_a|}2^{nR} 2^{-n(\Theta_\mathbf{s}-\delta(\epsilon))}
\yesnumber\label{eqn:PrErrEve2'_iid_noCSI}
\end{IEEEeqnarray*}
where (a) follows from \eqref{eqn:pmf_iid_noCSI} and $\mathbf{x}^{n}(j)[m]$ shows the $j$th element of vector $\mathbf{x}^{n}[m]$, (b) follows from joint typicality lemma and the memoryless property of the channel $\Xc\rightarrow\Yc$ according to \eqref{eqn:pmf_iid_noCSI}, (c) follows from \cite[Lemma~II.1]{Csis98}, \eqref{eqn:type3}, \eqref{eqn:type2}, where $\big(\!\binom{k}{n}\!\big)=\binom{n+k-1}{k-1}$ is the multiset number, (d) follows from the independence of disjoint paths, where we define
\begin{align*}
\Theta_\mathbf{s}=\inf\limits_{\substack{\prod\limits_{j=1}^{n_r}p_{j}(\mathbf{x}_{a}(j)|\mathbf{x}(j),\mathbf{s}(j))\\\forall j\in\{1,\ldots,n_r\}:E_{p_j}[d(\mathbf{X}_{a}(j),\mathbf{X}(j))]\leq D_{j,s}=s\cdot D_j}}\sum\limits_{j=1}^{n_r}I(\mathbf{X}(j);\mathbf{Y}_s(j))
\end{align*}

Therefore:
\begin{IEEEeqnarray*}{rcl}
Pr(\Ec_2)&\leq& \binom{n_r}{n_a}\max_{w_H(\mathbf{s})\leq n_a}Pr(\Ec'_{2,\mathbf{s}})\\
&\leq& \binom{n_r}{n_a}\max_{w_H(\mathbf{s})\leq n_a}n^{|\Xc|+|\Yc|+|\Xc_a|}2^{n(R-\Theta_\mathbf{s}+\delta(\epsilon))}\\
&=& \binom{n_r}{n_a}n^{|\Xc|+|\Yc|+|\Xc_a|}2^{n(R-\min\limits_{w_H(\mathbf{s})\leq n_a}\Theta_\mathbf{s}+\delta(\epsilon))}
\label{eqn:PrErrEve2_iid_noCSI}
\end{IEEEeqnarray*}
Hence, considering the finite alphabets, if $R\leq\min\limits_{\substack{\mathbf{s}\in\{0,1\}^{n_r}\\w_H(\mathbf{s})\leq n_a}}\Theta_\mathbf{s}-\delta(\epsilon)$, $Pr(\Ec_2)$ goes to zero as $n\rightarrow\infty$. This completes the achievablility proof.

\textit{Converse:} The converse easily follows from Fano's inequality, by noting that for \emph{every} $\mathbf{s}\in\{0,1\}^{n_r}:w_H(\mathbf{s})\leq n_a$ and every $\mathbf{h}^{n}$ (Definition~\ref{def:code}), we must have $H(M|\mathbf{Y}_s^n)\leq n\epsilon_n$ for some $\epsilon_n\stackrel{n\rightarrow\infty}{\longrightarrow} 0$.
\end{IEEEproof}

\begin{IEEEproof}[Proof of Theorem~\ref{thm:cap_iid_TxCSI}]
The proof is similar to that of Theorem~\ref{thm:cap_iid_noCSI}. Hence, we only describe the differences in the achievablility part.
Here, Tx generates $|\Sc|=\binom{n_r}{n_a}$ codebooks, $C_{\mathbf{s}}$, similar to the one in Theorem~\ref{thm:cap_iid_noCSI} (with fixed $p_{\mathbf{X}}(\mathbf{x})$ for each codebook). To transmit $m$, knowing the current state of the channel, $\mathbf{s}_i=\mathbf{s}$ for $i\in\{1,\ldots,n\}$, Tx selects $C_{\mathbf{s}}$ and transmits $\mathbf{x}^{n}[m]$ from that codebook. The rest of the proof is similar to Theorem~\ref{thm:cap_iid_noCSI}.
\end{IEEEproof}

\begin{IEEEproof}[Proof of Theorem~\ref{thm:low_mem_noCSI}]
We apply a random coding technique on top of a random linear code (Varshamov construction \cite{Var57}). To make this combination possible, we propose a new coding scheme by using proper auxiliary codewords.

First, consider the replacement attacks(defined in Section~\ref{subsec:main_niid}) and let $d_j=s\cdot2D_j,j\in[1:n_r]$.
Now, fix $p_{\mathbf{X}}(\mathbf{x})$ and generate $2^{nR}$ i.i.d sequences $\mathbf{u}^{k}[m]$ each with probability $\prod\limits_{i=1}^np_{\mathbf{X}}(\mathbf{u}_{i})$, for some $k\geq nR\log_{|\Xc|} 2$ where $m\in[1:2^{nR}]$.
Repeat the following codebook generation process $n_r$ times (for $j\in\{1,\ldots,n_r\}$) to produce $\mathbf{x}^{n}$:

Choose a random $|\Xc|$-ary matrix $\mathbf{G}_j\in\mathbb{F}_{|\Xc|}^{k\times n}$ whose elements are uniformly and independently chosen from $\mathbb{F}_{|\Xc|}$. Let $\Vc$ be the set of all $n$-length sequences in $\Xc^n$. Now, use the matrix $\mathbf{G}_j$ as a generator matrix to generate $2^{nR}$ sequences $\mathbf{x}^{n}(j)[m]=\mathbf{u}^{k}(j)[m]\mathbf{G}_j$ (Varshamov construction) with minimum distance $d_j$. Therefore, this code satisfies the Gilbert-Varshamov bound: for every $|\Xc|\geq2$ and real $0\leq d_j\leq 1-\frac{1}{|\Xc|}$, the volume of the Hamming ball centered at $\mathbf{x}^{n}(j)[m]$ ($\forall m\in[1:2^{nR}], j\in[1:n_r]$) is bounded as:
\begin{IEEEeqnarray*}{rcl}
|\Xc|^{nH_{|\Xc|}(d_j)-o(n)}\leq Vol_{|\Xc|}(\mathbf{x}^{n}(j)[m],n\times d_j)\leq |\Xc|^{nH_{|\Xc|}(d_j)}\yesnumber\label{eqn:vol_low_mem_noCSI}
\end{IEEEeqnarray*}

Then, pick the sequences in $\Vc$ that belong to these hamming balls and call them codewords, $\mathbf{v}^{n}(j)[m,l]$: $\mathbf{v}^{n}(j)[m,l]\in B_{|\Xc|}(\mathbf{x}^{n}(j)[m],n\times d_j)$, where $l$ shows each codeword's index in the ball, $l\in[1:Vol_{|\Xc|}(\mathbf{x}^{n}(j)[m],n\times d_j)]$. Let $L_j=\frac{1}{n\log_{|\Xc|}2}\max\limits_m \log_{|\Xc|}Vol_{|\Xc|}(\mathbf{x}^{n}(j)[m],n\times d_j)$. This means that the $\mathbf{v}^{n}(j)$ is selected according to $p(\mathbf{v}|\mathbf{x}):E_{p_j}[d(\mathbf{V}(j),\mathbf{X}(j))]\leq d_{j}$ for $j\in\{1,\ldots,n_r\}$.

To transmit $m$, Tx sends $\mathbf{x}^{n}[m]$. Rx after receiving $\mathbf{y}^n$, looks for a unique index $\tilde{m}$ and some $\tilde{l}$ such that:
\begin{IEEEeqnarray*}{c}
(\mathbf{v}^{n}[\tilde{m},\tilde{l}],\mathbf{y}^n)\in A_\epsilon^{n}(\mathbf{X}_a,\mathbf{Y}).
\end{IEEEeqnarray*}
Due to the symmetry of the random codebook generation, the probability of error is independent of the specific messages. Hence, to analyze the probability of error, without loss of generality, we assume that $m=1$ is encoded and transmitted. Note that although the foreseer adversaries' channel inputs are chosen with memory, the channel from the adversaries to the Rx is i.i.d, i.e.,
\begin{IEEEeqnarray*}{rcl}
p(\mathbf{y}^n|\mathbf{x}_a^n)&=&\prod\limits_{j=1}^{n_r}\prod\limits_{i=1}^{n}p_j(\mathbf{y}_i(j)|\mathbf{x}_{a,i}(j))\yesnumber\label{eqn:pmf_mem_noCSI}
\end{IEEEeqnarray*}
Due to \eqref{eqn:pmf3}, we have $\mathbf{x}_a^{n}(j)\in B_{|\Xc|}(\mathbf{x}^{n}(j)[1],n\times d_j)$ for $j\in[1:n_r]$. Thus, the error events at Rx are:
\begin{IEEEeqnarray*}{rcl}
\Ec_1&=&\{\forall\mathbf{s}\in\{0,1\}^{n_r}, w_H(\mathbf{s})\leq n_a, \nexists l':(\mathbf{v}^{n}[1,l'],\mathbf{y}^n)\in A_\epsilon^{n}(\mathbf{X}_a,\mathbf{Y})\}\\
\Ec_2&=&\{\exists\mathbf{s}\in\{0,1\}^{n_r}, w_H(\mathbf{s})\leq n_a, \text{ such that }(\mathbf{v}^{n}[m,l'],\mathbf{y}^n)\in A_\epsilon^{n}(\mathbf{X}_a,\mathbf{Y}) \text{ for some } m\neq 1\text{ and some } l'\}
\label{eqn:ErrEve_mem_noCSI}
\end{IEEEeqnarray*}

Based on the problem definition, we are sure that $\mathbf{x}_a^{n}(j)\in B_{|\Xc|}(\mathbf{x}^{n}(j)[m],n\times d_j)$. Since $\mathbf{v}^{n}(j)[m,l]$ covers all the codewords in this ball, $\mathbf{x}_a^{n}(j)=\mathbf{v}^{n}(j)[m,l']$ for some $l'$. Therefore, thanks to the AEP \cite{CovTho06}, $Pr\left(\Ec_1\right)\rightarrow 0$ as $n\rightarrow\infty$. Now, to consider the probability of $\Ec_2$, define:
\begin{IEEEeqnarray*}{rcl}
\!\!\Ec'_{2,\mathbf{s}}&=&\{\exists m\neq 1 \text{such that} (\mathbf{v}^{n}[m,l'],\mathbf{y}^n)\in A_\epsilon^{n}(\mathbf{X}_a,\mathbf{Y}) \text{for some\,} l'\}
\end{IEEEeqnarray*}
for every $\mathbf{s}\in\{0,1\}^{n_r}$. Considering \eqref{eqn:pmf_mem_noCSI}, the joint AEP \cite{CovTho06} implies:
\begin{IEEEeqnarray*}{l}
Pr(\Ec'_{2,\mathbf{s}})\leq 2^{n(R+\sum\limits_{j=1}^{n_r}L_j)}2^{-n(H(\mathbf{Y})+H(\mathbf{V})-H(\mathbf{X}_a,\mathbf{Y})-\epsilon)}
\end{IEEEeqnarray*}
Therefore, if $R+\sum\limits_{j=1}^{n_r}L_j\leq H(\mathbf{V})-H(\mathbf{X}_a|\mathbf{Y})-\delta(\epsilon)$, $Pr(\Ec'_{2,\mathbf{s}})$ goes to zero as $n\rightarrow\infty$. Using \eqref{eqn:vol_low_mem_noCSI} and the disjoint path property, we have:
\begin{IEEEeqnarray}{l}
R\leq \sum\limits_{j=1}^{n_r}[H(\mathbf{V}(j))-H(\mathbf{X}_a(j)|\mathbf{Y}(j))-\frac{H_{|\Xc|}(d_j)}{\log_{|\Xc|} 2} ]-\delta(\epsilon)\qquad
\label{eqn:Rs_mem_noCSI}
\end{IEEEeqnarray}
for all $\mathbf{h}^n$. Thus, $Pr(\Ec_2)\leq \binom{n_r}{n_a}\max_{w_H(\mathbf{s})\leq n_a}Pr(\Ec'_{2,\mathbf{s}})$ goes to zero if \eqref{eqn:Rs_mem_noCSI} holds for all $\mathbf{s}\in\{0,1\}^{n_r}:w_H(\mathbf{s})\leq n_a$ which results in $R\leq \min\limits_{\substack{\mathbf{s}\in\{0,1\}^{n_r}\\w_H(\mathbf{s})\leq n_a}}\inf\limits_{\mathbf{h}^n}\sum\limits_{j=1}^{n_r}[H(\mathbf{V}(j))-H(\mathbf{X}_a(j)|\mathbf{Y}(j))-\frac{H_{|\Xc|}(d_j)}{\log_{|\Xc|} 2}]-\delta(\epsilon)$.
The proof for erasing attacks is similar by defining $d_j=s\cdot D_j$ and noting that a code with minimum distance $d_j$ can recover from $d_j$ erasures. This completes the proof.
\end{IEEEproof}

\begin{IEEEproof}[Proof of Theorem~\ref{thm:low_mem_TxCSI}]
The proof is straightforward considering the proofs of Theorems~\ref{thm:cap_iid_TxCSI} and \ref{thm:low_mem_noCSI}.
\end{IEEEproof}
\begin{IEEEproof}[Proof of Theorem~\ref{thm:up_mem}]

\textit{No CSI}: We use the asymptotic Hamming bound (i.e., sphere packing bound) to limit the rate of a code that wishes to correct $D_j$ errors. Similar to the proof of Theorem~\ref{thm:low_mem_noCSI}, first consider the replacement attacks with $d_j=s\cdot2D_j,j\in[1:n_r]$. As $M\rightarrow \mathbf{X}(j)\rightarrow \mathbf{X}_a(j) \rightarrow \mathbf{Y}(j)$ forms a Markov chain for $j\in\{1,\ldots,n_r\}$, using Fano's inequality, for every $\mathbf{h}^{n},h^{n}(j):\Xc^{n}\times\{0,1\}\mapsto\Xc_a^{n}$  satisfying \eqref{eqn:pmf3}, we have:
\begin{IEEEeqnarray*}{rcl}
H(\mathbf{X}^n_a|\mathbf{Y}^n)\leq H(M|\mathbf{Y}^n)\leq n\epsilon_n
\label{eqn:fano1_mem_noCSI}
\end{IEEEeqnarray*}
where $\epsilon_n\rightarrow0$ as $n\rightarrow\infty$. Hence
\begin{IEEEeqnarray}{rcl}
H(\mathbf{X}_a^n)&\leq& I(\mathbf{X}_a^n;\mathbf{Y}^n)+n\epsilon_n\nonumber\\
&\stackrel{(a)}{\leq}& n\sum\limits_{j=1}^{n_r}I(\mathbf{X}_a(j);\mathbf{Y}(j))+n\epsilon_n
\label{eqn:fano2_mem_noCSI}
\end{IEEEeqnarray}
for all $\mathbf{h}^n$, where (a) follows from \eqref{eqn:pmf}.

To consider all possible $\mathbf{h}^{n}$, we must consider the type class $\Tc^n(\Pc^n_{j,s}(\Xc\times\Yc))$. Recall that one must be able to correct any possible $\frac{d_j}{2}=s\cdot D_j$ errors made by the adversary on $j$th route. This means that the minimum distance of the codewords must be greater than $d_j$. Otherwise, the adversary intentionally always chooses the closer codeword that cannot be distinguished at the Rx. Thus, the rate of the code for \emph{every} $\mathbf{s}\in\{0,1\}^{n_r}:w_H(\mathbf{s})\leq n_a$ and every $\mathbf{h}^{n}$ (Definition~\ref{def:code}) must satisfy the Hamming bound (for a minimum distance $d_j$):
\begin{IEEEeqnarray*}{rcl}
2^{nR}&\stackrel{(a)}{\leq}&\frac{|\Tc^n_{\textbf{s}}(\Pc^n(\Xc\times\Yc))|}{\prod\limits_{j=1}^{n_r}\sum_{l=0}^{\frac{d_j}{2}}\binom{l}{n}(|\Xc|-1)^l}\\
&\leq&\frac{|\Tc^n_{\textbf{s}}(\Pc^n(\Xc\times\Yc))|}{\prod\limits_{j=1}^{n_r}|\Xc|^{nH_{|\Xc|}(\frac{d_j}{2})}}\\
&\stackrel{(b)}{\leq}&\frac{2^{nH(\mathbf{X}^n_a)}}{\prod\limits_{j=1}^{n_r}|\Xc|^{nH_{|\Xc|}(\frac{d_j}{2})}}
\stackrel{(c)}{\leq}\prod\limits_{j=1}^{n_r}\frac{2^{nI(\mathbf{X}_a(j);\mathbf{Y}(j))+n\epsilon_n}}{|\Xc|^{nH_{|\Xc|}(\frac{d_j}{2})}}\label{eqn:up_mem_ham}
\end{IEEEeqnarray*}
for all $\mathbf{h}^n$, where (a) follows by defining $\Tc^n_{\textbf{s}}(\Pc^n(\Xc\times\Yc))=\{(\textbf{x}^n,\textbf{y}^n):(\textbf{x}^n(j),\textbf{y}^n(j))\in\Tc^n_{j,s}(\Pc^n_{j,s}(\Xc\times\Yc))\}$, (b) follows from \cite[Lemma~II.2]{Csis98}, (c) follows from \eqref{eqn:fano2_mem_noCSI}.
For erasing attacks, it is enough to define $d_j=s\cdot D_j$ and note that a code with minimum distance $d_j$ can recover from $d_j$ erasures. Therefore, \eqref{eqn:up_mem_noCSI} is proved.
The proof for the case of CSI at Tx follows a similar lines by considering the proof of Theorem~\ref{thm:cap_iid_TxCSI}. This completes the proof.
\end{IEEEproof}

\begin{IEEEproof}[Proof of Corollary~\ref{corr:cap_iid_BEC}]
Let $P_j=Pr(\mathbf{X}(j)=1)$ and without loss of generality assume that $P_j\leq \frac{1}{2}$. Also, let $Pr(\mathbf{X}_a(j)=e)=N'_j$; considering the distortion measure defined above with finite distortion limits (i.e., $D_j$s) and adversaries' model in Definition \ref{def:code}, we have $N'_j=Pr(\mathbf{X}_a(j)\neq\mathbf{X}(j))\leq \mathbf{s}(j)\cdot D_j$. Thus,
\begin{IEEEeqnarray*}{rcl}
H(\mathbf{Y}_s(j))&=&H((1-P_j)(1-N'_j)(1-N_j),P_j(1-N'_j)(1-N_j),1-(1-N'_j)(1-N_j))\\
&=&H((1-N'_j)(1-N_j))+(1-N'_j)(1-N_j)H(P_j)\\
H(\mathbf{Y}_s(j)|\mathbf{X}(j))&=&H((1-N'_j)(1-N_j))
\end{IEEEeqnarray*}
Therefore, we have: $I(\mathbf{X}(j);\mathbf{Y}_s(j))=(1-N'_j)(1-N_j)H(P_j)$. Now, we can obtain \eqref{eqn:cap_iid_noCSI}, as:
\begin{IEEEeqnarray*}{l}
\Cc_{i.i.d}^{\text{nC}}=
\sup_{0 \leq D_j \leq P_j\leq \frac{1}{2}}\min_{\substack{\mathbf{s}\in\{0,1\}^{n_r}\\w_H(\mathbf{s})\leq n_a}}\;\sum\limits_{j=1}^{n_r}
\min_{N'_j\leq \mathbf{s}(j)\cdot D_j}(1-N'_j)(1-N_j)H(P_j).
\end{IEEEeqnarray*}
which will be maximized for $P_j=\frac{1}{2}$ independent of $\mathbf{s}(j)$ for $j\in\{1,\ldots,n_r\}$. Hence, the rest of the proof is straightforward.
\end{IEEEproof}

\begin{IEEEproof}[Proof of Corollary~\ref{corr:low_mem_BEC}]
Let $\mathbf{V}_i(j)=\mathbf{X}_{i}(j)$ for $j\in\{1,\ldots,n_r\}$, $P_j=Pr(\mathbf{X}(j)=1)$ and $N'_j\doteq\mathbf{s}(j)\cdot D_j$.
Recall that for all $\mathbf{h}^n$ (Definition~\ref{def:code}, satisfying \eqref{eqn:pmf3}), we have $Pr(\mathbf{X}(j)\neq\mathbf{X}_a(j))=Pr(\mathbf{X}_a(j)=e)\leq N'_j$. After some calculations, one can compute:
\begin{IEEEeqnarray*}{l}
H(\mathbf{V}(j))=H(P_j)\\
H(\mathbf{X}_a(j)|\mathbf{Y}(j))\leq \bar{N}_jH(\frac{N'_j}{\bar{N}_j})+N_j(1-N'_j)H(P_j)
\end{IEEEeqnarray*}
where we defined $\bar{N}_j=N_j(1-N'_j)+N'_j$.
The rest of the proof is straightforward.
\end{IEEEproof}

\begin{IEEEproof}[Proof of Corollary~\ref{corr:up_mem_BEC}]
Similarly to the proof of Corollary~\ref{corr:up_mem_BSC}, it is enough to choose the proper joint distribution for the adversaries' input. Hence, let $\mathbf{X}_{a,i}(j)=\text{BEC}(\mathbf{X}_{i}(j),\mathbf{S}(j)\cdot D_j)$. Computing the mutual information term in \eqref{eqn:up_mem_noCSI} completes the proof.
\end{IEEEproof}

\begin{IEEEproof}[Proof of Corollary~\ref{corr:cap_iid_Gaus}]
The achievablility follows by the standard arguments that extend the achievable rate to the Gaussian case with continuous alphabets \cite{CovTho06}. As we assume Gaussian channel inputs, let $\mathbf{X}_a(j)\sim \Nc(0,P_{a,j})$ where $E[d(\mathbf{X}_{a}^n(j),\mathbf{X}^n(j))]\leq \mathbf{s}(j)\cdot D_j$. To find the $\inf\sum\limits_{j=1}^{n_r}I(\mathbf{X}(j);\mathbf{Y}_s(j))$ in \eqref{eqn:cap_iid_noCSI}, first we find a lower bound and then we show it is achievable by the adversaries.
\begin{IEEEeqnarray*}{rcl}
I(\mathbf{X}(j);\mathbf{Y}_s(j))&=&H(\mathbf{Y}_s(j))-H(\mathbf{Y}_s(j)|\mathbf{X}(j))\yesnumber\label{eqn:Gaus_iid_ent}\\
&\geq& H(\mathbf{X}_a(j)+\mathbf{Z}(j))-H(\mathbf{X}_a(j)-\mathbf{X}(j)+\mathbf{Z}(j))\\
&=& \frac{1}{2}\log(2\pi e (P_{a,j}+N_j))-H(\mathbf{X}(j)-\mathbf{X}_a(j)-\mathbf{Z}(j))\\
&\stackrel{(a)}{\geq}& \frac{1}{2}\log(2\pi e (P_{a,j}+N_j))- \frac{1}{2}\log(2\pi e (\mathbf{s}(j)\cdot D_j+N_j))
\end{IEEEeqnarray*}
where in (a) we use $E[d(\mathbf{X}_{a}^n(j),\mathbf{X}^n(j))]\leq \mathbf{s}(j)\cdot D_j$. \eqref{eqn:Gaus_iid_ent} should be minimized over all $P_{a,j}$ satisfying the distortion limit. Thus, $P_{a,j}=P_j-\mathbf{s}(j)\cdot D_j$. This lower bound is achievable by adversaries if they choose a joint distribution with a backward Gaussian test channel, $\mathbf{X}_{i}(j)=\mathbf{X}_{a,i}(j)+\mathbf{Z'}_i(j)$, where $\mathbf{Z'}_i(j)\sim \Nc(0,\mathbf{s}(j)\cdot D_j)$ and $\mathbf{X}_{a,i}(j)\sim \Nc(0,P_j-\mathbf{s}(j)\cdot D_j)$.
The rest of the proof is straightforward.
\end{IEEEproof}

\end{document}